\documentclass[10pt,conference]{IEEEtran}
\IEEEoverridecommandlockouts

\usepackage{cite}
\usepackage{amsmath,amssymb,amsfonts}
\usepackage{algorithmic}
\usepackage{graphicx}
\usepackage{textcomp}
\usepackage{xcolor}
\usepackage{multirow}
\usepackage{booktabs}
\usepackage{url}

\usepackage[framemethod=tikz]{mdframed}
\mdfdefinestyle{mpdframe}{
    frametitlebackgroundcolor   =black!15,
    frametitlerule              =true,
    roundcorner                 =5pt,
    middlelinewidth             =1pt,
    innermargin                 =0.3cm,
    outermargin                 =0.3cm,
    innerleftmargin             =0.3cm,
    innerrightmargin            =0.3cm,
    innertopmargin              =0.5cm,
    innerbottommargin           =0.5cm
}

\usepackage{xspace}

\newcommand{\eg}{\emph{e.g.,}\xspace}

\newcommand{\approach}{\textit{ISM}\xspace}

\newcommand{\approachSIM}{\textit{ISM}\xspace}

\newcommand{\baselineSIM}{TechTube\xspace}

\usepackage[numbers, square,sort&compress]{natbib}
\usepackage{enumitem}
\setlist[itemize]{noitemsep, topsep=0pt, leftmargin=*}
\usepackage{listings}
\usepackage{caption}
\usepackage{xcolor}
\definecolor{verylightgray}{rgb}{0.95, 0.95, 0.95}

\begin{document}

\title{Intelligent Semantic Matching (ISM) for Video Tutorial Search using Transformer Models\thanks{Author accepted manuscript. Published in the 2025 IEEE/ACM 22nd International Conference on Mining Software Repositories (MSR), pp. 712--724, 2025. DOI: \protect\url{https://doi.org/10.1109/MSR66628.2025.00108}.\newline
\copyright\ 2025 IEEE. Personal use of this material is permitted. Permission from IEEE must be obtained for all other uses, in any current or future media, including reprinting/republishing this material for advertising or promotional purposes, creating new collective works, for resale or redistribution to servers or lists, or reuse of any copyrighted component of this work in other works.}}

\author{\IEEEauthorblockN{Ahmad J. Tayeb}
\IEEEauthorblockA{
King Abdulaziz University, Saudi Arabia \\
ajtayeb@kau.edu.sa}
\and
\IEEEauthorblockN{Sonia Haiduc}
\IEEEauthorblockA{
Florida State University, United States \\
shaiduc@fsu.edu
}}

\maketitle

\begin{abstract}
The rise in the number and diversity of available software development video tutorials has enhanced digital learning for developers but also introduced challenges in locating relevant content efficiently. Existing video search methods, including keyword-based approaches and tools like CodeTube and TechTube, rely primarily on retrieval algorithms such as BM25, which fail to capture the semantic nuances and user intentions behind search queries. To address these limitations, we introduce \approachSIM, an approach that uses SBERT to generate semantically rich vectors from video tutorial transcripts to improve the search for programming video tutorials. By segmenting transcripts and implementing a re-ranking process, \approachSIM effectively preserves context and enhances the relevance of search results. Additionally, \approachSIM generates informative video summaries using GPT-4, allowing developers to quickly assess the relevance of video content. To evaluate our approach, we first performed a quantitative study comparing \approachSIM with the baseline \baselineSIM. The results revealed that \approachSIM performs better in both video retrieval and fragment identification, achieving a Hit@5 score of 0.95 and an average F1 score of 0.70 compared to the baseline’s 0.58 and 0.52, respectively. We also performed a user study, which revealed that users strongly preferred the semantic matching capabilities and AI-generated summaries of our approach. This work advances the state-of-the-art in programming video tutorial search and summarization by offering more nuanced and user-aligned retrieval and summarization mechanisms.

\end{abstract}

\section{Introduction}\label{sec:introduction}

Over time, software development video tutorials have become a popular learning resource among computer science students and professional developers \cite{escobar2019survey}, with millions of videos available on many programming topics. Even as the era of LLMs begins to unfold, video tutorials remain one of the preferred resources for learning programming concepts and solving programming issues \cite{tayeb2024icsme}. However, the growth in popularity and availability of this type of resources has also introduced notable challenges in discovering relevant and specific content in a sea of available information. As the volume of these tutorials increases, it becomes increasingly challenging to sift through them and find pertinent information.

Researchers have explored methods to enhance the search, discovery, navigation, and usefulness of programming video tutorials by extracting and indexing information implicitly captured in them, such as code snippets, GUIs, or code elements \cite{ponzanelli2016too,ponzanelli2017automatic, khandwala2018codemotion, bao2020psc2code, bao2015reverse,bao2017extracting,ott2018deep,malkadi2023improving, alahmadi2020ui}, or  by identifying key fragments that represent the video content \cite{ponzanelli2016too,ponzanelli2017automatic, vahedi2021summarizing}.
Despite this progress, current video search engines mainly rely on keyword-based methods, like Apache Lucene’s BM25, which use a bag-of-words model and struggle to capture semantic meaning and user intent \cite{cichosz2018case}.

Vahedi et al. \cite{vahedi2021summarizing} aimed to address this issue with their approach, \baselineSIM, which uses pseudo-relevance feedback to expand developers' video search queries with frequently used keywords from the audio transcript, in the hopes of capturing more user intent and context. However, this approach has its limitations as it can overlook contextually relevant videos that use unusual terminology. Also, frequently used keywords may not always represent the video's full context \cite{jiang2014necessary}. Moreover, \baselineSIM depends on the structure and wording of developer queries, potentially missing nuanced needs, especially in complex queries \cite{karisani2016query,yan2017pseudo}. While \baselineSIM marks progress over basic keyword-based search, its limitations underscore the need for advanced methods that capture and interpret both video content and user intent.

To address these gaps, we introduce our approach \textit{Intelligent Semantic Matching} (\approachSIM), which leverages Transformer-based models to create semantically dense vectors from video transcripts. Although Transformer models like BERT effectively capture context, they are limited by input size, often around 512 tokens \cite{devlin2018bert,reimers2019sentence}, making their direct application to lengthy video transcripts impractical. Segmenting transcripts into smaller chunks fits the input constraints but risks losing broader context across segments.
To address these challenges, our solution involves re-ranking these segmented vectors by relevance, preserving continuity across multiple relevant chunks from the same video. This re-ranking process manages cases where several chunks from a single video are relevant, ensuring broader context alignment with the developer’s query. Additionally, we integrate GPT-4 in our approach to generate summaries of the video content, helping learners quickly assess the relevance of videos without watching them in full.

We evaluated \approachSIM by comparing it with \baselineSIM on the \baselineSIM dataset \cite{vahedi2021summarizing}. Our results show that \approachSIM significantly outperforms \baselineSIM in video retrieval, achieving a Hit@5 score of 0.95 compared to 0.58 for \baselineSIM, and in identifying relevant video fragments with an F1 score of 0.70, exceeding \baselineSIM's 0.52.
In addition, we conducted a user study to compare \approachSIM's effectiveness and user satisfaction with \baselineSIM's. We also evaluated user preferences between our GPT-4-generated video summaries and the video descriptions provided by the original creators. Findings showed that users favored \approachSIM, noting its semantic matching capabilities, and preferred the GPT-4-generated summaries, which they found accurate and informative.

Below is a summary of our contributions:

\begin{itemize}
    \item We propose \approachSIM, an approach that leverages Transformer-based models to create semantically dense vectors from video tutorial transcripts, overcoming the limitations of keyword-based search engines and enhancing programming video search and retrieval capabilities.
    \item We empirically evaluated our approach against the baseline \baselineSIM, demonstrating improvements in video retrieval and relevant video fragment identification.
    \item We conducted a user study to assess the effectiveness and user satisfaction of \approachSIM compared to the baseline \baselineSIM, as well as user preferences between video content summarizing methods (\eg original video descriptions by video creators vs. GPT-4-generated summaries).
    \item We provide a replication package that includes the source code of our approach, enabling the reproducibility of our results. The package also contains the materials and data used in the user study, facilitating further research and comparison \cite{ism_replicationPackage}.
\end{itemize}

\section{Related Work}\label{sec:Related Work}

Software development video tutorials are vital for knowledge sharing among developers, allowing them to document and disseminate expertise while building their online presence \cite{macleod2015code, macleod2017documenting}. Programming screencasts have been shown to be useful in addressing specific developer queries about API usage, database operations, system setup, game development, etc. \cite{ellmann2017find}. However, the increasing number of available  videos presents challenges in efficiently finding relevant information.

Various tools and approaches have been developed to extract information embedded in video tutorials and enhance their searchability. ACE uses Optical Character Recognition (OCR) to extract code fragments from video frames and then consolidates code snippets across frames using statistical language models \cite{yadid2016extracting}. Other approaches used CNN-based methods in combination with OCR to accurately identify and extract code from video frames \cite{alahmadi2018accurately, alahmadi2020code} or recognize and extract typeset and handwritten code \cite{ott2018deep, ott2018learning}. PSC2Code employs a CNN-based classifier to denoise video frames and applies OCR to extract source code, improving code retrieval precision \cite{bao2020psc2code}. CodeTube also uses OCR to extract code fragments from video frames, segmenting videos into coherent fragments classified by purpose, such as theoretical concepts or code implementation \cite{ponzanelli2016too, ponzanelli2017automatic}. TechTube segments video content by analyzing silent periods in the audio, effectively identifying and summarizing relevant video segments, and outperforming CodeTube in retrieving relevant search results \cite{vahedi2021summarizing}.

Other relevant tools analyzing or extracting information from programming screencasts include PSFinder, which identifies live-coding screencasts by detecting IDE windows in video frames, achieving high precision in distinguishing such videos \cite{yang2022efficient}. Sieve uses transfer learning and word embeddings to filter and extract software-related content from platforms like Twitter and YouTube \cite{sulistya2020sieve}. Codemotion expands learner interactions with video programming tutorials by making source code within videos interactive, allowing for inline code editing and in-video exercises \cite{khandwala2018codemotion}. ScvRipper, on the other hand, automates the extraction of time-series HCI data from videos \cite{bao2017extracting}. Other approaches have focused on detecting and extracting Java code and GUI elements \cite{alahmadi2022vid2meta}, XML data \cite{alahmadi2022vid2xml}, or UI screens \cite{alahmadi2020ui} from Android screencasts.

In terms of action recognition, tools like ActionNet and SeeHow identify and extract developer actions in screencasts \cite{zhao2019actionnet, zhao2023seehow}. ActionNet detects actions like code editing using image differencing and CNNs, while SeeHow analyzes code-line edits to present workflows, making screencasts more intuitive \cite{zhao2023seehow}.

A few studies have evaluated or focused on improving OCR engines' performance in extracting code from video tutorials. One empirical study compared various OCR engines for extracting code from screencasts and provided guidelines on optimizing factors like font and image resolution \cite{malkadi2020study}. Then, CodeT5-OCRfix enhanced code extraction accuracy by fine-tuning the CodeT5 model using OCRed and ground truth code pairs \cite{malkadi2023improving}. Similarly, another study has also shown that enhancements in OCR performance for code extraction from videos can be achieved using image super-resolution and large language models \cite{alahmadi2024optimizing}.

Previous research has also focused on classifying comments that users post in response to programming video tutorials on YouTube, with the goal of distinguishing relevant from irrelevant comments \cite{poche2017analyzing}. Other work has used various approaches to tag video tutorials based on their content and therefore facilitate the quick identification of relevant tutorials \cite{parra2018automatic}.

Other prior work focused on linking programming screencast content with other types of resources or software artifacts. For example, FVT connects video fragments to text tutorials by extracting code snippets as anchors \cite{nong2019fvt}. The work by Moleshi et al. linked video tutorials to source code artifacts by mining spoken words and GUI text \cite{moslehi2018feature, moslehi2020feature}.

Our work builds on these efforts by introducing \approach, which utilizes Transformer-based models \cite{vaswani2017attention} to generate dense vectors from video transcripts. Unlike previous methods like CodeTube and TechTube, which rely on keyword retrieval methods such as BM25 that lack semantic understanding, our approach uses SBERT embeddings and cosine similarity to encode documents and queries separately, reducing computational overhead while maintaining high accuracy \cite{reimers2019sentence}. We segment long transcripts into manageable chunks and aggregate results, bypassing traditional methods like BM25. We also incorporate LLMs for summarization, leveraging their zero-shot capabilities to produce summaries \cite{pu2023summarization}.

\section{Methodology}\label{sec:Methodology}
\begin{figure*}[]
\centering
\includegraphics[width=0.8\linewidth]{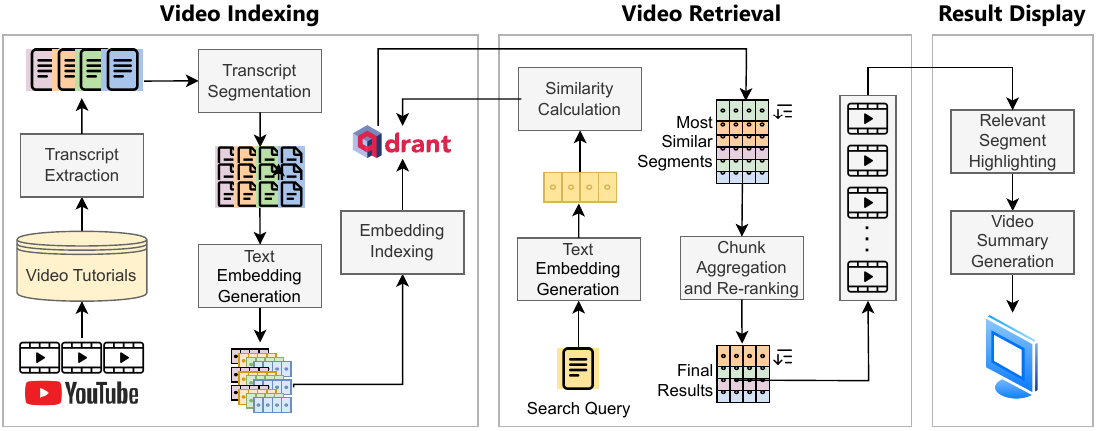}
\caption{The overview of our approach, \approachSIM.}
\label{fig:semantic_matching:approach_overview}
\end{figure*}

Our approach, \approachSIM, is designed to enhance the search and retrieval of video tutorials by leveraging Transformer-based models to encode the semantic content of video transcripts into dense vector representations. These vector representations are then compared using cosine similarity, aggregated, and ranked to ensure that the most relevant videos are prioritized in response to user queries. \approachSIM\ comprises three main components: Video Indexing, Video Retrieval, and Result Display. An overview of these components is illustrated in Figure~\ref{fig:semantic_matching:approach_overview}.

\subsection{Video Indexing}

The Video Indexing component is responsible for preparing and structuring the data from video tutorials to facilitate efficient search and retrieval. It begins by utilizing the YouTube Data API to download metadata associated with video tutorials, such as video IDs, titles, and descriptions. Subsequently, the video transcripts (\eg, captions) are downloaded. These transcripts can be obtained directly from YouTube, which stores either the manual transcription provided by the video creator or a transcript generated by YouTube's automatic speech recognition. In cases where a transcript is unavailable (\eg, if the creator has disabled it), we extract the video's audio track and transcribe it using Whisper\footnote{https://openai.com/index/whisper/}, a state-of-the-art speech recognition model. The resulting transcript, whether obtained from YouTube or Whisper, includes timestamp information, which is critical for aligning text with the corresponding video segments during query processing.

After obtaining the transcript, we segment it into smaller chunks to comply with the input length constraints of Transformer models. We empirically determined that chunk sizes of 80 words yielded the best retrieval performance. For each chunk, we generate an embedding vector using the \textit{"multi-qa-distilbert-cos-v1"} model, a pre-trained Sentence-BERT (S-BERT) model, which captures the semantic content of the transcript chunks. We specifically chose this model for video search due to its optimization for question-answering tasks. This model is derived from DistilBERT, a smaller and more efficient version of BERT \cite{sanh2019distilbert}, making it well-suited for efficiently matching user queries with relevant video content.

These embedding vectors, along with the associated metadata, are indexed using Qdrant\footnote{https://qdrant.tech/}, an open-source and scalable vector search engine. Qdrant's support for multi-dimensional vector search makes it an ideal choice for our application, enabling efficient retrieval of relevant video chunks during query processing.

\subsection{Video Retrieval}\label{sec:semantic_matching:Approach:Video Retrieval}

The Video Retrieval component processes user queries and runs a search to identify the most relevant video tutorial chunks to the query. When a user submits a textual query, the system processes it using the same S-BERT model employed for transcript chunk embeddings, generating a query embedding vector. This query embedding is then used to retrieve the top-$k$ most relevant video chunks from the indexed dataset in Qdrant by calculating the cosine similarity between the query vector and the video chunk vectors.

The cosine similarity between the query embedding \( q \) and each chunk embedding \( c_j \) is computed as follows:

\begin{equation*}
\text{Sim}(q, c_j) = \frac{q \cdot c_j}{\|q\| \|c_j\|}
\end{equation*}

where \( q \) is the vector representation of the query, \( c_j \) is the vector representation of the \( j \)-th chunk, \( \cdot \) denotes the dot product, and \( \| \cdot \| \) represents the Euclidean norm of a vector.

To balance the coverage of relevant content and minimize the impact of videos with an excessive number of chunks, we aggregate video chunks originating from the same video and re-rank the videos based on the relevance scores of their chunks.

For all strategies, except the Maximum Similarity Score, we consider a subset of chunks per video that has retrieved chunks, defined as:
\begin{equation*}
n(V_i) = \min(N_i, n_{\text{max}})
\end{equation*}
where \( N_i \) is the total number of retrieved chunks for video \( V_i \), and \( n_{\text{max}} \) is the maximum number of top chunks to consider per video (e.g., 4). The top \( n \) chunks are selected based on their similarity scores. This approach focuses on the most relevant \( n \) chunks, preventing videos with many chunks from disproportionately influencing the final relevance score or rank.

The re-ranking strategies are defined as follows:

\begin{itemize}
    \item \textbf{Maximum Similarity Score (MaxSim):} This strategy considers only the single chunk with the highest similarity score for each video \( V_i \) that has retrieved chunks. This approach highlights the most relevant segment of each video:
    \begin{equation*}
    \text{Score}(V_i) = \max_{c_j \in C_i} \{\text{Sim}(q, c_j)\}
    \end{equation*}
    where \( C_i \) is the set of all retrieved chunks for video \( V_i \). Videos are sorted in descending order based on \( \text{Score}(V_i) \).

    \item \textbf{Average Rank Aggregation (AvgRank):} This strategy calculates the final rank for a video \( V_i \) by averaging the ranks of its top \( n \) relevant chunks. Lower average ranks indicate higher relevance:
    \begin{equation*}
    \text{Rank}(V_i) = \frac{1}{n} \sum_{j=1}^{n} \text{Rank}(c_j)
    \end{equation*}
    where \( \text{Rank}(c_j) \) is the original rank position of chunk \( c_j \) in the retrieved results. Videos are then sorted in ascending order based on \( \text{Rank}(V_i) \).

    \item \textbf{Average Score Aggregation (AvgScore):} This strategy computes the final score for a video \( V_i \) by averaging the relevance scores of its top \( n \) relevant chunks:
    \begin{equation*}
    \text{Score}(V_i) = \frac{1}{n} \sum_{j=1}^{n} \text{Sim}(q, c_j)
    \end{equation*}
    where \( \text{Sim}(q, c_j) \) represents the cosine similarity between the query embedding \( q \) and chunk embedding \( c_j \). Videos are sorted in descending order based on \( \text{Score}(V_i) \).

    \item \textbf{Weighted Rank Adjustment (WeightRank):} In this strategy, each chunk's rank contributes to the final video rank inversely proportional to its position, giving more weight to higher-ranked chunks:
    \begin{equation*}
    \text{Rank}(V_i) = \sum_{j=1}^{n} \frac{\text{Rank}(c_j)}{j}
    \end{equation*}
    Lower \( j \) values (i.e., chunks appearing earlier in the retrieval list) have greater influence. Videos are sorted in ascending order based on \( \text{Rank}(V_i) \).

    \item \textbf{Weighted Score Adjustment (WeightScore):} This strategy enhances the influence of higher-ranked chunks by weighting their scores inversely proportional to their position:
    \begin{equation*}
    \text{Score}(V_i) = \sum_{j=1}^{n} \frac{\text{Sim}(q, c_j)}{j}
    \end{equation*}
    Here, earlier chunks (with lower \( j \)) contribute more significantly to the final score. Videos are sorted in descending order based on \( \text{Score}(V_i) \).
\end{itemize}

Based on our empirical evaluation, detailed in Section \ref{sec:Empirical Evaluation}, the \textit{Weighted Score Adjustment} strategy provided the best overall results. This approach balances the influence of individual chunks, prioritizing videos that contain highly relevant segments while ensuring that the presence of a large number of chunks does not disproportionately affect the final ranking.

\subsection{Result Display}

The Result Display component presents the final search results to the user in an intuitive and informative manner. It comprises two main submodules: Relevant Segment Highlighting and Video Summary Generation.

\subsubsection{Relevant Segment Highlighting}

This submodule is designed to identify and display the most relevant video segments in response to a user's query, allowing users to quickly access the most relevant portions of a video without viewing the entire content. The process begins with calculating the cosine similarity between the user's query embedding and each chunk embedding within the video transcript, as previously described. For this step, we use a different S-BERT model (\textit{"all-mpnet-base-v2"}) and segment size (80 words) to better align with the focus on identifying precise, relevant segments. Since a word can consist of one or more tokens, this segmentation approach helps keep each chunk within the model's token limit while capturing the local context. We chose this model because of its superior performance in semantic similarity tasks \cite{song2020mpnet}. Its larger, more complex architecture, combined with training on a diverse dataset, allows it to capture deeper semantic meanings in text \cite{song2020mpnet}, making it effective for identifying contextually relevant video segments.

Once the similarity scores are computed, the submodule filters the chunks by applying a similarity threshold. Specifically, it retains only those chunks whose similarity scores meet or exceed a certain percentage \( \alpha \) (e.g., 45\%) of the maximum similarity score observed across all chunks:

\begin{equation*}
\text{Sim}_{\max} = \max_{c_j \in C_i}\{ \text{Sim}(q, c_j)\}
\end{equation*}
\begin{equation*}
C_i' = \{c_j \in C_i \mid \text{Sim}(q, c_j) \geq \alpha \cdot \text{Sim}_{\max}\}
\end{equation*}

Here, \( C_i \) is the set of all chunks in video \( V_i \), \( C_i' \) is the subset of chunks that meet the similarity threshold, and \( \text{Sim}_{\max}\) represents the maximum similarity score among all chunks. This thresholding ensures that only the most relevant chunks, relative to the query, are considered for further analysis.

\begin{figure}[h]
\centering
\includegraphics[width=1.05\linewidth]{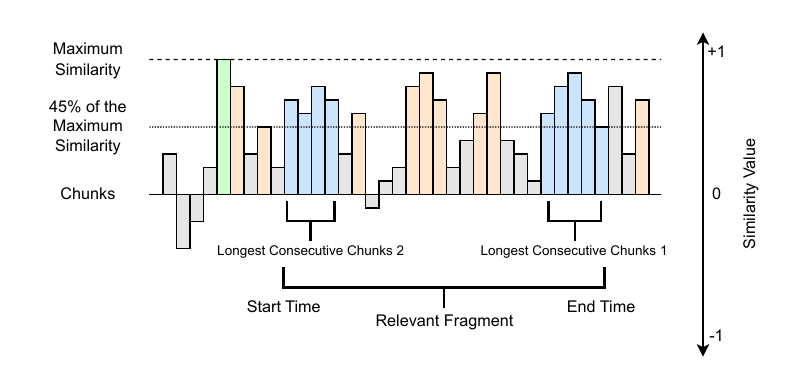}
\caption{Overview of the Relevant Segment Identification Process in \approachSIM.}
\label{fig:semantic_matching:relevant_fragments}
\end{figure}

The filtered chunks are then analyzed to identify sequences of consecutive chunks, which are grouped together if they appear next to each other in the video transcript without interruption. These consecutive chunks are prioritized based on their length, with the submodule focusing particularly on the two longest sequences, as shown in Figure~\ref{fig:semantic_matching:relevant_fragments}. Longer sequences are more likely to capture a complete and contextually relevant segment of the video.

To provide additional context, the start and end times of the selected segments are adjusted by subtracting a few seconds from the start time and adding a few seconds to the end time. This helps capture the full scope of the relevant content, allowing the user to understand the segment in context.

The final output of the submodule is a set of video segments that are most relevant to the user's query, with their adjusted start and end times ensuring that important contextual information is included. This process significantly enhances the efficiency and effectiveness of the search and retrieval process by presenting users with the most relevant and contextually complete portions of the video tutorial.

\subsubsection{Video Summary Generation}

This submodule leverages GPT-4 to generate concise, natural language summaries of video content. The video transcript is split into segments that adhere to GPT-4's input length constraints, and a summary is generated for each segment using the prompt shown in Listing \ref{lst:chunk_summarization_prompt}. Each segment summary is limited to 150 words to ensure conciseness and user readability.

\vspace{0.1cm}

\begin{lstlisting}[caption={Prompt used for summarizing video transcript chunks}, label=lst:chunk_summarization_prompt]
Summarize the following chunk of a video tutorial transcript in a concise manner, focusing on the key points and ensuring the summary is complete:\n\n[CHUNK_TEXT]
\end{lstlisting}

\vspace{0.2cm}

If the transcript is split into more than one chunk, resulting in multiple segment summaries, GPT-4 is further prompted to condense these summaries into a single, coherent summary of the entire video. This final summary is generated using the prompt shown in Listing \ref{lst:video_summarization_prompt}.

\vspace{0.1cm}

\begin{lstlisting}[caption={Prompt used for summarizing a video tutorial by combining chunk summaries}, label=lst:video_summarization_prompt]
Summarize the video tutorial through the following combined summaries of a video transcript in a concise manner, focusing on the key points and ensuring the summary is complete:\n\n [COMBINED_CHUNK_SUMMARIES]
\end{lstlisting}

\vspace{0.2cm}

The Result Display component ultimately presents the re-ranked videos, their relevant segments, and the GPT4-generated summaries in a user-friendly interface. This allows users to quickly understand the key content of the videos and identify the most relevant tutorials for their needs.

\section{Empirical Evaluation}\label{sec:Empirical Evaluation}
In this section, we describe the empirical evaluation of our approach, \approachSIM, addressing two research questions (RQs): (RQ1) How does \approachSIM\ perform in retrieving relevant technical videos in response to queries? and (RQ2) How effective is \approachSIM\ in identifying the most relevant sections of these videos?

\subsection{Baseline: TechTube}
In our empirical evaluation, we compare the performance of \approachSIM\ against TechTube \cite{vahedi2021summarizing}, the closest related and best performing approach to date, which serves as our baseline. TechTube is designed to assist developers in finding relevant technical videos and identifying the most pertinent segments within those videos.

\paragraph{Retrieving Relevant Videos} First, TechTube retrieves relevant videos based on a natural language query given as input. The system then enhances this query using pseudo-relevance feedback based on Rocchio’s expansion technique. This technique involves initially retrieving a set of most relevant videos to the original query. From these initial results, TechTube selects the topmost videos and extracts additional keywords that are closely associated with the query. These additional keywords are used to reformulate and expand the original query, making it more comprehensive and better aligned with the user’s intent. The enhanced query is then matched against a database of preprocessed technical videos, retrieving the most relevant videos related to the reformulated query. TechTube uses Apache Lucene\footnote{https://lucene.apache.org/} as its search engine.

\paragraph{Retrieving Relevant Video Fragments} To retrieve relevant video fragments, TechTube first segments the video into smaller chunks by detecting silences in the audio track. These silences often indicate shifts in topics or significant pauses, which serve as natural boundaries for segmenting the content. Each chunk's content is then transcribed using speech recognition software. The system measures the similarity between the user’s query and these transcriptions to identify the most relevant video fragments. Once the relevant chunks are identified, TechTube highlights and combines them using techniques such as Longest Common Subsequences (LCS) to form a video summary that addresses the user’s query. The summary is basically a shorter video clip that represents part of the original video.

The authors of TechTube \cite{vahedi2021summarizing} have demonstrated that their approach outperforms CodeTube \cite{ponzanelli2016too}, an earlier approach that enhances search precision by using Optical Character Recognition (OCR) to identify and retrieve code fragments within video tutorials. This further establishes TechTube as a robust baseline for evaluating the performance of \approachSIM. By comparing our method against TechTube, we aim to demonstrate the effectiveness and improvements offered by \approachSIM\ in both video retrieval and fragment identification.

\subsection{Dataset}

We used the dataset from the TechTube paper \cite{vahedi2021summarizing} for our empirical evaluation. This dataset includes 98 natural language queries related to online repository maintenance (e.g., GitHub) and programming tasks in Java and Python, sourced from a benchmark and top-rated Stack Overflow threads. Based on the search results of the queries, the authors downloaded a set of 400 YouTube videos, each over three minutes long with English audio. From these, they selected 98 videos, each matching a query.

The TechTube replication package provided the initial and reformulated queries, which we used to evaluate \approachSIM\ in RQ1, concerning relevant video retrieval. It also contained ground truth data of relevant video fragments, which we utilized in RQ2 to assess \approachSIM's success in retrieving relevant fragments.

To replicate TechTube's search functionality, we used Lucene, more specifically Apache Solr\footnote{https://solr.apache.org}, which implements Lucene's search functionality as a service, and used the provided \textit{reformulated} queries and ground truth to compare TechTube's results with those of \approachSIM on the same dataset.

Twelve videos were unavailable on YouTube at the time of our study, reducing the dataset to 86 videos. For RQ1, we excluded one query lacking a reformulated version, leaving 85 queries. For RQ2, we excluded two queries due to processing issues in TechTube's package, resulting in 84 queries for evaluating video fragment retrieval.

\subsection{Evaluation Metrics}
To assess the performance of \approachSIM\ comparing with the baseline, we use several standard evaluation metrics. Specifically, Hit@k and Mean Reciprocal Rank (MRR) are used to evaluate the retrieval of relevant videos, while Precision, Recall, and F1-Score are used to evaluate the retrieval of relevant video fragments. Below, we provide the definitions for each metric.

\paragraph{\textbf{Hit@k}} A metric that checks whether at least one relevant video appears in the top \( k \) search results. The overall Hit@k metric is then computed as:

\begin{equation*}
\text{Hit@k}(Q) = \frac{\sum_{q \in Q} \text{hit@k}(q)}{|Q|}
\end{equation*}

where \( Q \) is the set of all queries. This metric is used to evaluate the retrieval of relevant videos.

\paragraph{\textbf{Mean Reciprocal Rank (MRR):}} The average of the reciprocal ranks of results for queries and is defined as:

\begin{equation*}
\text{MRR}(Q) = \frac{1}{|Q|} \sum_{q \in Q} \frac{1}{\text{rank}(q, K)}
\end{equation*}

MRR captures how early the first relevant video appears in the results and is used to evaluate the retrieval of relevant videos.

\paragraph{\textbf{Precision}} In the context of video segment retrieval, it measures the proportion of the predicted segment that correctly overlaps with the ground truth segment.

\paragraph{\textbf{Recall}} It measures the proportion of the ground truth segment that is correctly retrieved by the predicted segment.

\paragraph{\textbf{F1-Score}} It is the harmonic mean of precision and recall, providing a balanced measure that considers both false positives and false negatives.

\subsection{RQ1: Video Retrieval Performance}
To address RQ1, we evaluated the performance of \approachSIM\ for retrieving relevant technical videos in response to textual queries. We compared \approachSIM\ using five different strategies (AvgRank, AvgScore, MaxSim, WeightRank, and WeightScore) against two baselines: \textbf{Lucene}, which simply searches based on the \textit{original} queries, and \textbf{TechTube}, which uses Lucene with the \textit{reformulated} queries from the replication package.

\begin{figure}[]
\centering
\begin{minipage}{0.67\linewidth}
    \centering
    \includegraphics[width=\linewidth]{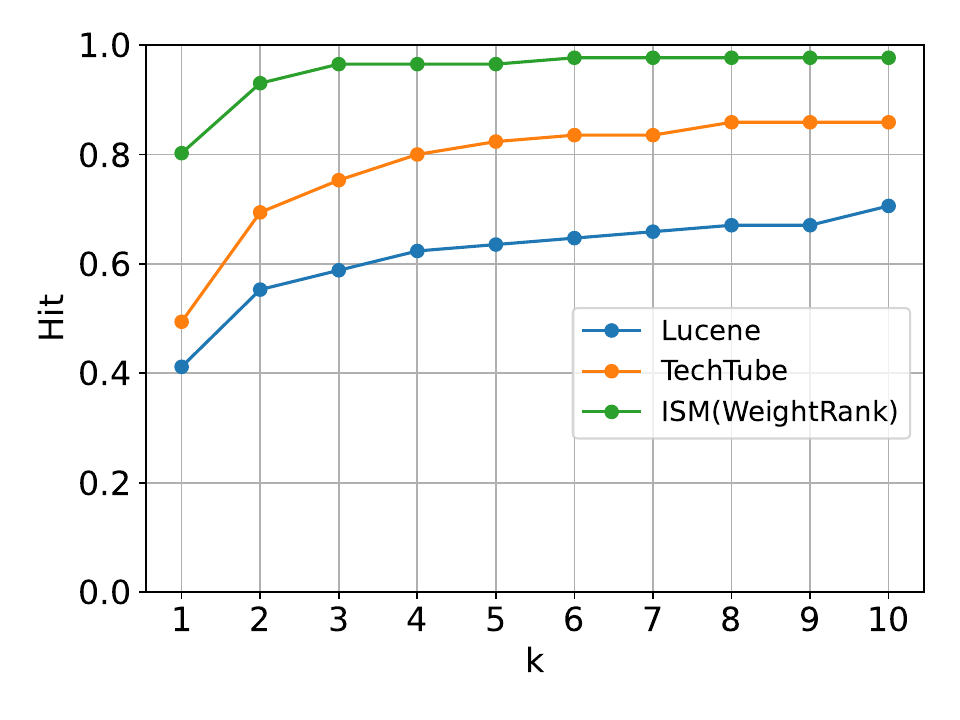}
\end{minipage}
\hfill
\begin{minipage}{0.67\linewidth}
    \centering
    \includegraphics[width=\linewidth]{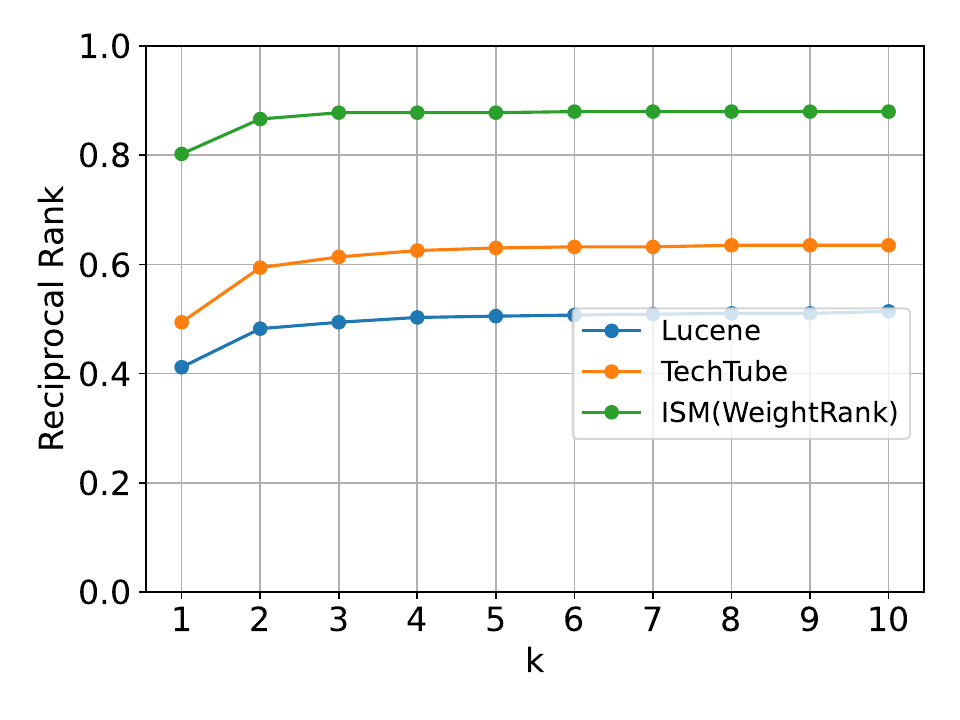}
\end{minipage}
\caption{Hit@k and Mean Reciprocal Rank (MRR) results across various \( k \) values.}
\label{fig:semantic_matching:Hit_and_MRR}
\end{figure}

\begin{table*}[h]
\centering
\caption{Hit@k and Mean Reciprocal Rank (MRR) for Video Retrieval}
\label{tab:hit_mrr_combined}
\resizebox{\textwidth}{!}{

\begin{tabular}{lcccccccccc||cccccccccc}
\toprule
\multirow{2}{*}{Method} & \multicolumn{10}{c}{Hit@k} & \multicolumn{10}{c}{MRR@k} \\
\cmidrule(lr){2-11} \cmidrule(lr){12-21}
& 1 & 2 & 3 & 4 & 5 & 6 & 7 & 8 & 9 & 10 & 1 & 2 & 3 & 4 & 5 & 6 & 7 & 8 & 9 & 10 \\
\midrule
Lucene           & 0.41  & 0.55  & 0.59  & 0.62  & 0.64  & 0.65  & 0.66  & 0.67  & 0.67  & 0.71   & 0.41  & 0.48  & 0.49  & 0.50  & 0.51  & 0.51  & 0.51  & 0.51  & 0.51  & 0.51   \\
TechTube         & 0.49  & 0.69  & 0.75  & 0.80  & 0.82  & 0.84  & 0.84  & 0.86  & 0.86  & 0.86   & 0.49  & 0.59  & 0.61  & 0.63  & 0.63  & 0.63  & 0.63  & 0.64  & 0.64  & 0.64   \\
\midrule
ISM(AvgRank)     & 0.77  & 0.92  & 0.95  & 0.95  & \textbf{0.98}  & \textbf{0.98}  & \textbf{0.98}  & \textbf{0.98}  & \textbf{0.98}  & \textbf{0.98}   & 0.77  & 0.84  & 0.85  & 0.85  & 0.86  & 0.86  & 0.86  & 0.86  & 0.86  & 0.86   \\
ISM(AvgScore)    & 0.79  & 0.93  & 0.95  & \textbf{0.97}  & 0.97  & \textbf{0.98}  & \textbf{0.98}  & \textbf{0.98}  & \textbf{0.98}  & \textbf{0.98}   & 0.79  & 0.86  & 0.87  & 0.87  & 0.87  & 0.87  & 0.87  & 0.87  & 0.87  & 0.87   \\
ISM(MaxSim)      & \textbf{0.80}  & 0.86  & 0.95  & 0.95  & 0.97  & 0.97  & \textbf{0.98}  & \textbf{0.98}  & \textbf{0.98}  & \textbf{0.98}   & \textbf{0.80} & 0.83  & 0.86  & 0.86  & 0.86  & 0.86  & 0.87  & 0.87  & 0.87  & 0.87   \\
ISM(WeightRank)  & \textbf{0.80}  & 0.93  & \textbf{0.97}  & \textbf{0.97}  & 0.97  & \textbf{0.98}  & \textbf{0.98}  & \textbf{0.98}  & \textbf{0.98}  & \textbf{0.98}   & \textbf{0.80} & \textbf{0.87} & \textbf{0.88} & \textbf{0.88} & \textbf{0.88} & \textbf{0.88} & \textbf{0.88} & \textbf{0.88} & \textbf{0.88} & \textbf{0.88} \\
ISM(WeightScore) & 0.78  & \textbf{0.94}  & \textbf{0.97}  & \textbf{0.97}  & 0.97  & \textbf{0.98}  & \textbf{0.98}  & \textbf{0.98}  & \textbf{0.98}  & \textbf{0.98}   & 0.78  & 0.86  & 0.87  & 0.87  & 0.87  & 0.87  & 0.87  & 0.87  & 0.87  & 0.87 \\
\bottomrule
\end{tabular}
}
\end{table*}

Table \ref{tab:hit_mrr_combined} shows the Hit@k results across various \( k \) values. The results demonstrate that \approachSIM\ outperformed both Lucene and TechTube across all strategies. For instance, at \( k=5 \), TechTube achieved a Hit@5 of 0.82, while \approachSIM\ achieved up to 0.98 across several strategies, such as AvgRank, AvgScore, MaxSim, and WeightRank. Notably, the Weighted Score Adjustment (WeightRank) strategy of \approachSIM\ consistently performed well, achieving the highest Hit@1 of 0.80 and maintaining a high retrieval performance across \( k=10 \), with all strategies reaching 0.98 for Hit@10.
\begin{minipage}{0.5\textwidth}
    \centering

\end{minipage}
\begin{minipage}{0.5\textwidth}
    \centering

\end{minipage}

We conducted statistical significance tests on the retrieval performance to further validate these improvements. First, we performed Shapiro-Wilk tests \cite{shapiro1965analysis} to assess the normality of the data, which indicated that the distributions for both \approachSIM\ and the baseline (TechTube) were not normally distributed (p-values < 0.05). Given this, we employed the Wilcoxon signed-rank test—a non-parametric test that does not assume normality and is commonly used for non-normally distributed data \cite{wilcoxon1992individual}. The results revealed a significant difference between \approachSIM\ and the baseline (Wilcoxon: p-value = 0.0126), confirming that the observed improvements in Hit@10 are statistically significant.

As described in Figure~\ref{fig:semantic_matching:Hit_and_MRR}, this improvement in performance highlights the advantage of using Transformer-based models to generate semantically dense vector representations of video transcripts, which enhances the ability to retrieve relevant videos. Furthermore, by aggregating video chunks and applying advanced re-ranking strategies like Weighted Score Adjustment, \approachSIM\ was able to prioritize more contextually relevant videos, improving the overall retrieval performance.

Additionally, the results in Table \ref{tab:hit_mrr_combined} show the Mean Reciprocal Rank (MRR) for video retrieval. Similar to the Hit@k results, \approachSIM\ consistently outperformed both baselines across all \( k \) values. For example, the WeightRank strategy achieved the highest MRR at 0.87 when considering the top 2 videos, whereas TechTube achieved only 0.59. This further confirms that \approachSIM\ not only retrieves the relevant videos more frequently but also ranks them higher, allowing users to find the most relevant video faster.

\subsection{RQ2: Video Segment Retrieval Performance}
\begin{table}[th]
\centering
\caption{Video Segment Retrieval Performance}
\begin{tabular}{lccc}
\toprule
Method & Average Precision & Average Recall & Average F-Score \\
\midrule
TechTube & 0.67 & 0.55 & 0.52 \\
ISM & \textbf{0.69} & \textbf{0.82} & \textbf{0.70} \\
\bottomrule
\end{tabular}
\label{tab:segment_retrieval}
\end{table}

For RQ2, we evaluated the effectiveness of \approachSIM\ in identifying the most relevant segments from the retrieved videos. We compared the segment retrieval performance of \approachSIM\ against TechTube in terms of Average Precision, Average Recall, and F1-Score.

\approachSIM\ demonstrated superior performance in segment retrieval as well, as shown in Table~\ref{tab:segment_retrieval}. The Average Precision of \approachSIM\ was 0.69, slightly higher than TechTube’s 0.67. However, \approachSIM\ showed a substantial improvement in recall, achieving 0.82 compared to TechTube’s 0.55. This significant improvement in recall indicates that \approachSIM\ is better at capturing more relevant segments within the videos.

\begin{figure}[h]
\centering
\includegraphics[width=0.95\linewidth]{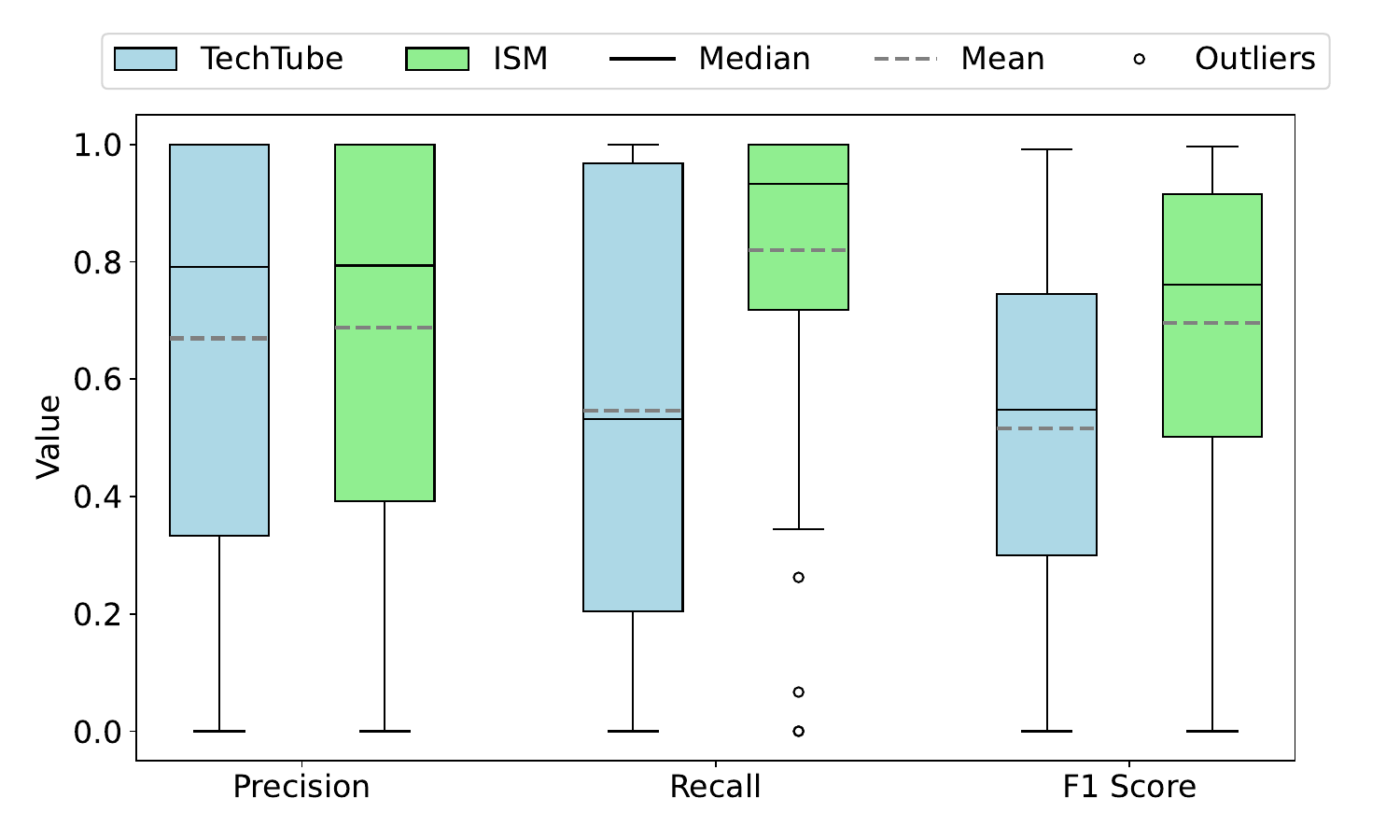}
\caption{Comparison of segment retrieval metrics between \approachSIM\ and TechTube.}
\label{fig:semantic_matching:boxplot_metrics_comparison}
\end{figure}

As shown in Figure~\ref{fig:semantic_matching:boxplot_metrics_comparison}, the improved recall resulted in a higher F1-Score for \approachSIM\ (0.70), outperforming TechTube’s F1-Score of 0.52. This demonstrates that \approachSIM\ not only retrieves more relevant video segments but does so with a greater balance between precision and recall, making it more effective for users who need to identify relevant sections of technical videos to their queries.

Furthermore, statistical significance tests confirmed these improvements. The Wilcoxon signed-rank test showed a statistically significant difference in F1-Scores between \approachSIM\ and TechTube (Wilcoxon: p-value = 1.39e-05), indicating that \approachSIM\ provides a measurable advantage in segment retrieval performance.

These results underline the efficacy of \approachSIM\ in both retrieving relevant videos and precisely identifying the key segments that match users' informational needs. The combination of advanced semantic representations and tailored re-ranking strategies allows \approachSIM\ to outperform traditional keyword-based search engines like Lucene and state-of-the-art baselines like TechTube.

\section{User Study}\label{sec:User Study}
We conducted a user study to evaluate the effectiveness and user satisfaction with \approachSIM\ compared to the baseline (TechTube) in more realistic usage scenarios, where users can write their own queries. The study also aimed to assess different \textit{text summaries for videos}, specifically comparing the original video descriptions provided by the video creators with the GPT-4-generated summaries. The study received ethics approval from our Institutional Review Board (IRB), and we adhered to all protocols to ensure ethical compliance.

\subsection{Study Design}\label{sec:semantic_matching:User_Study:Study_Design}
The study was divided into three phases, where participants engaged in search tasks, expressed their preference between video descriptions by video creators and GPT-4-generated summaries, and rated the quality of GPT-4-generated summaries. We implemented a web application where we hosted our study, and participants could perform all the study tasks from within the web tool.

\paragraph{Phase 1: Video Search Evaluation}
Participants were assigned short Python programming tasks and used two different search engines inside our tool—ISM and TechTube—to find relevant video tutorials. Participants were first asked to input their search queries into a text input field within the tool. In response, two lists of search results were displayed side by side: one generated by ISM and the other by TechTube. Participants were asked to compare the results from both lists and evaluate which search engine provided more helpful results for their queries on a scale from 1 to 4, where 1 was the lowest rating and 4 was the highest. Additionally, they were asked to check the checkbox below every video in the result lists that they found relevant to their query. To minimize bias, the source of the search results (ISM or TechTube) was hidden, and the placement of the two search result lists was randomly switched between tasks.

\paragraph{Phase 2: Textual Summary Preference Evaluation}
Participants were shown a set of random videos with two summaries provided below them. Their task was to rate each summary based on how well it described the video content in general. Participants were asked to rate each summary on a scale from 1 to 4, where 1 was the lowest rating and 4 was the highest. After rating the summaries, they were asked to select the summary they preferred or liked the most. Participants were instructed to watch the video to get an idea of its content before rating the summaries.

Participants were also informed that if the video did not appear for some reason, or if they had difficulty understanding the video content, or if one of the summaries only contained links or tags without any description of the video, they should skip the video by clicking the "Skip" button and get another video instead. Similar to Phase 1, the source of the summaries was hidden, and the placement of the summaries was randomly switched between evaluations to prevent bias.

\paragraph{Phase 3: Quality of GPT-4-Generated Summaries}
Participants were asked to watch a set of random short videos accompanied by a summary of the video’s content for each. They were tasked with rating the summary based on two criteria—accuracy and completeness—on a scale from 1 to 4, where 1 was the lowest rating and 4 was the highest. Participants were also given the option to provide additional feedback on the summary in a text box provided.

To clarify the evaluation criteria, participants were informed that accuracy refers to the extent to which the summary correctly reflects the information presented in the video, while completeness refers to the extent to which the summary captures all key information presented in the video. It was also explained that a summary could be highly accurate but omit important details (lacking completeness), while a complete summary might include all key information but misrepresent or incorrectly state some details (lacking accuracy). If participants encountered any issues with the video (such as it not displaying for some reason), they were instructed to skip the video and select another by clicking the "Skip" button.

In each phase, participants were asked to provide reasons for their choices (e.g., why they preferred a specific search engine or a particular type of summary). This additional feedback helped us gain insights into the reasoning behind their preferences.

To provide diverse programming tasks for the study, we collected 32 Python exercises from GeeksforGeeks\footnote{https://www.geeksforgeeks.org/python-exercises-practice-questions-and-solutions/}, covering a wide range of topics. Python was chosen due to its popularity among programming students and novices, which represent the main target of our study, due to the likelihood of them watching video tutorials for guidance \cite{tayeb2024icsme}. For getting the video data for the study, we first used the titles of the GeeksforGeeks tasks as queries to fetch relevant videos from YouTube via its API. A total of 640 videos were retrieved (20 videos per task). After filtering out clearly irrelevant videos, 583 videos were indexed for use in the search engine evaluation.

Participation in the study was voluntary and anonymous. Participants were randomly assigned tasks and videos for each phase, and in all phases, the search results and summaries produced by the different approaches were presented in random order to avoid positional bias. Two versions of the user study were designed: a short version lasting 15-20 minutes and a longer version lasting 30-40 minutes, depending on the number of tasks and videos presented. Participants were generally offered the longer version, but just in a few cases where people wanted to participate but expressed a lack of time, we offered the shorter version of the study.

\subsection{Participants}\label{sec:semantic_matching:User_Study:Participants}
A total of 91 participants completed the study, with the majority being undergraduate computer science students. Of the 91 participants, 3 chose to take the short version of the user study. Participants were asked to provide demographic information, including gender, programming experience, and current occupation.

In terms of gender, 67\% of the participants were male, 30\% were female, 2\% identified as non-binary, and 1\% preferred not to disclose their gender. Regarding general programming experience, 42\% of participants had 1-2 years of experience, 48\% had 2-5 years of experience, and 10\% had over 5 years of programming experience.

Focusing on Python programming experience, 57\% of participants reported having less than 1 year of experience, 12\% had 1-2 years, and 20\% had no experience in Python at all. Although some of our participants did not have experience programming in Python, the tasks we chose are simple enough to resemble tasks that someone with a little programming experience in other languages could understand and potentially implement in Python with the help of tutorials. In terms of occupation, the vast majority—93\%—were undergraduate students, while 1\% were graduate students and 5\% were software developers.

\subsection{Results of Phase 1: Search Engine Comparison}\label{sec:semantic_matching:User_Study:Results_Phase1}
In the first phase of the study, participants used \approachSIM and TechTube to complete Python programming tasks. They formulated queries based on the task descriptions provided and used the two search engines to get relevant videos. After conducting their searches, they rated the relevance of the results and chose the search engine they found more helpful.

\begin{figure}[h]
\centering
\includegraphics[width=0.7\linewidth]{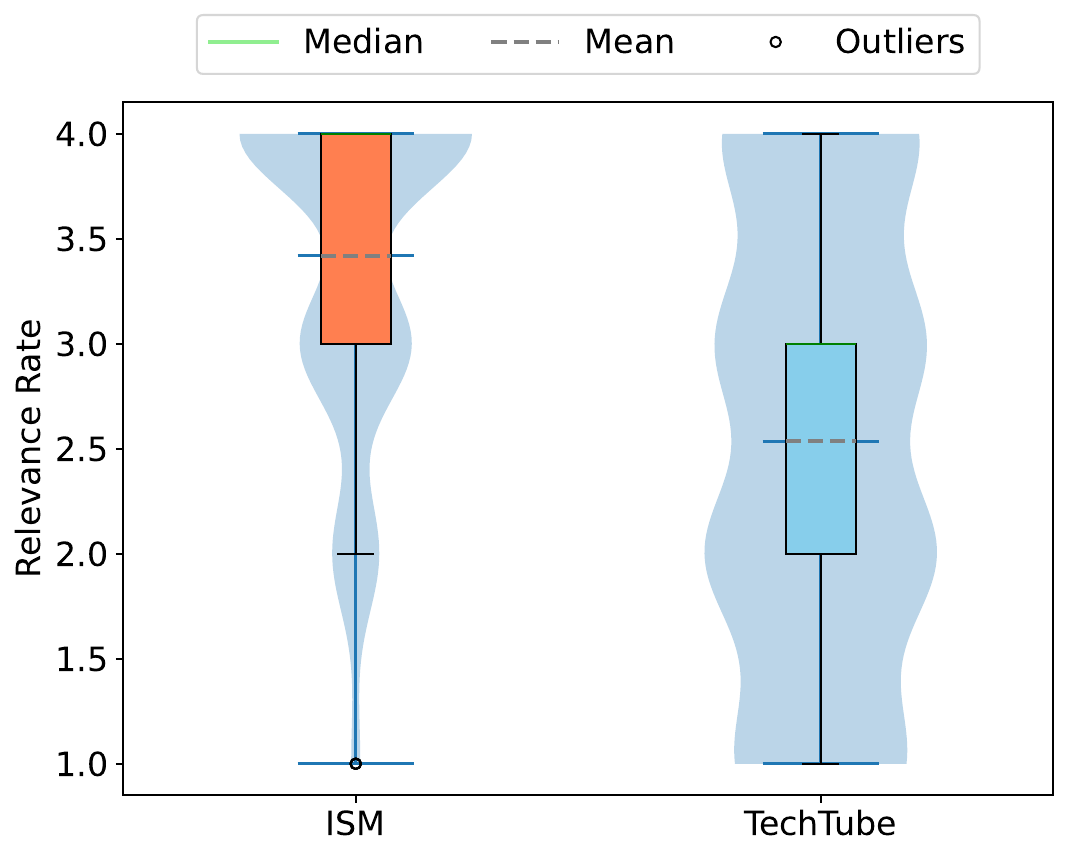}

\caption{Comparison of participant ratings for search result relevance between \approachSIM and TechTube during Phase 1 of the search task. The violin plot shows the distribution of relevance scores, while the overlaid boxplot highlights key statistics such as the median, mean, and interquartile range.}

\label{fig:semantic_matching:user_study_phase1_results_search_relevance_mixplot}
\end{figure}

Out of 179 searches conducted, in 64\% of the cases, participants preferred \approachSIM’s search results, while only in 20\% of cases they favored TechTube. Additionally, 11\% of the time participants rated both engines as equally useful, and only in 5\% of cases they found neither engine helpful. In terms of search result relevance, \approachSIM scored an average of 3.42 out of 4 (85\%), while TechTube scored 2.54 out of 4 (63\%). As shown in Figure~\ref{fig:semantic_matching:user_study_phase1_results_search_relevance_mixplot}, \approachSIM displays a higher median relevance score and a tighter interquartile range compared to TechTube, indicating more consistent relevance ratings for \approachSIM.

We recorded the number of queries participants made to complete their tasks. Most participants (76\%) were able to complete their tasks with just one query (136 queries in total), highlighting \approachSIM's efficiency in retrieving relevant content. A smaller percentage (14\%) needed two queries, while only 6\% needed three. The remaining participants required between four and twelve queries to complete their tasks, with a few participants (2\%) requiring more than five queries.

\subsubsection*{\textbf{Analysis of User-Selected Relevant Results Based on Checkbox Feedback}}

In addition to rating the search engines, participants were asked to check the checkbox below every video in the result lists that they found relevant to their query. We analyzed 179 search queries, but not all queries had relevance data.We filtered out 91 cases where participants did not mark any videos as relevant in either search engine's list. This could have been due to the optional nature of relevance marking—participants were not required to check every result—or because they did not find the results relevant.

This left us with 171 queries where participants actively marked at least one video as relevant. For these queries, we calculated the relevance rates for both \approachSIM (ISM) and TechTube. The relevance rate is defined as the proportion of results in each search list that were marked relevant by the participants. To ensure consistency and avoid accidental oversight, if a video appeared in both search lists and was marked as relevant in one list, the system automatically marked it as relevant in the other list as well.

For \approachSIM, the average relevance rate was 0.587, meaning that participants found nearly 59\% of the videos shown in \approachSIM’s search results to be relevant. The median relevance rate for \approachSIM was 0.50. In contrast, TechTube had a lower average relevance rate of 0.27, meaning only about 27\% of its search results were marked as relevant. The median relevance rate for TechTube was 0.20, further highlighting the gap in relevance between the two search engines.

By filtering out queries where no relevance was marked, we ensured that the relevance rates accurately reflect the participants' active engagement with the search results. This focused analysis highlights the clear advantage of \approachSIM (ISM) over TechTube in delivering relevant content. With \approachSIM achieving an average relevance rate of 59\%, compared to just 27\% for TechTube, the findings strongly support the conclusion that participants found \approachSIM’s search results significantly more useful and relevant to their tasks. This further reinforces the overall preference for \approachSIM in both relevance and participant satisfaction during the video search task.

In terms of video overlap between the two search engines, there were 209 videos that appeared in both \approachSIM and TechTube’s result lists across all 262 search queries. This overlap accounted for 17.46\% of the total 1,197 results presented by \approachSIM and 16.18\% of the 1,292 results presented by TechTube. Although the two engines returned a portion of similar content, the clear differences in relevance rates highlight \approachSIM’s superior ability to deliver relevant results based on user feedback.

\subsubsection*{\textbf{Analysis of User Feedback}}

Out of 104 responses, most users favored ISM for its relevant and specific search results, often positioned at the top. A few preferred TechTube for broader options, though both platforms had similar feedback on video quality. Some users noted both engines provided useful results, while a few expressed dissatisfaction with relevance on either platform.

\subsection{Results of Phase 2: Summary Preference Evaluation}\label{sec:semantic_matching:User_Study:Results_Phase2}
In phase two, participants were asked to evaluate and compare two types of summaries for each video—one written by the video creators (\eg video descriptions) and another generated by the GPT-4 model. They rated each piece of textual information based on how well it described the video’s content and then selected their preferred text.

\begin{figure}[h]
\centering
\includegraphics[width=0.7\linewidth]{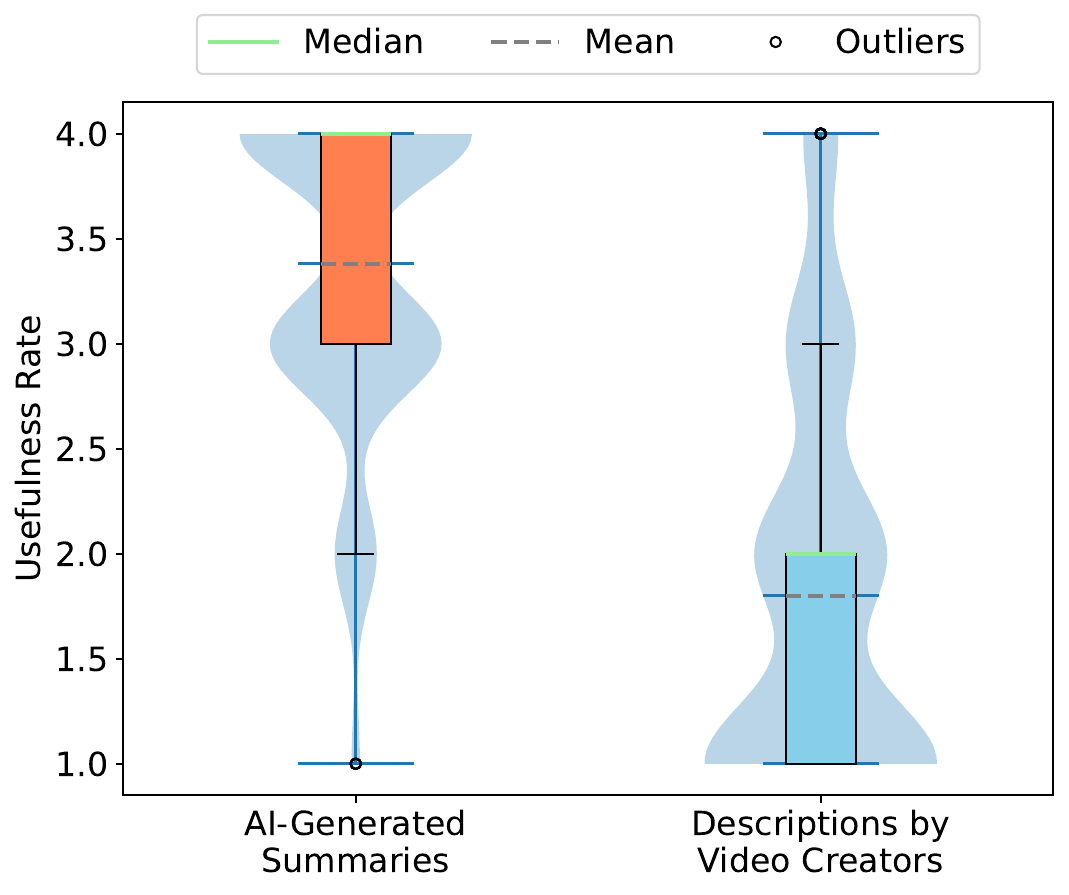}

\caption{Participant ratings of the usefulness of GPT-4-generated summaries versus video creator descriptions (Phase 2). The violin plot illustrates the distribution of rating scores, with overlaid boxplots showing key metrics, including the median, mean, and interquartile range.}
\vspace{-0.2cm}

\label{fig:semantic_matching:user_study_phase2_results_text_usefulness_mixplot}
\end{figure}

Out of 270 evaluations, 84\% chose the GPT-4-generated summaries, while only 11\% preferred the descriptions written by video creators. A small percentage (5\%) found both or neither useful. When asked to rate the usefulness of the textual information, \approachSIM’s AI-generated summaries scored 3.38 out of 4 (85\%), significantly higher than the video creators' descriptions, which scored 1.8 out of 4 (45\%). As shown in Figure~\ref{fig:semantic_matching:user_study_phase2_results_text_usefulness_mixplot}, GPT-4-generated summaries have a higher median rating score and a tighter interquartile range compared to video creator descriptions, indicating a more favorable and consistent perception of the AI-generated summaries.

\subsubsection*{\textbf{Analysis of User Feedback}}
An analysis of the 270 participant responses revealed a strong preference for the GPT-4-generated summaries over original video descriptions. Many participants criticized the video descriptions for lacking adequate information, often merely restating the title, resembling transcripts, or including irrelevant content like links and advertisements. In contrast, the GPT-4 summaries were praised for providing detailed and clear overviews of the video's key points, capturing the overall ideas more effectively, and being more readable and organized.

While a smaller group favored the original video descriptions for their conciseness, directness, and better formatting, some participants noted that the AI summaries could be too long or wordy. A few respondents felt that both the summaries and descriptions were equally good or served different purposes, and a minority believed that neither text provided a satisfactory explanation of the video content.

\subsection{Results of Phase 3: Quality of GPT-4 Summaries}\label{sec:semantic_matching:User_Study:Results_Phase3}
In the third phase, participants were asked to watch or skim random short videos accompanied by their summaries of the videos' contents. They were tasked with rating each summary based on two criteria—accuracy and completeness—on a scale from 1 to 4, where 1 was the lowest rating and 4 was the highest. Participants were also given the option to provide additional feedback on the summary in a text box provided.

The GPT-4-generated summaries received high ratings for both accuracy and completeness. Participants gave an average accuracy score of 3.39 out of 4 (85\%) and an average completeness score of 3.08 out of 4 (77\%). These results indicate that the GPT-4-generated summaries were effective in providing a concise and accurate overview of the video content, allowing users to quickly grasp the key points and determine whether the video was useful for their task at hand without needing to watch the entire video.

\subsubsection*{\textbf{Analysis of User Feedback}}
Based on 69 participant responses, several key themes emerged regarding the quality of the AI-generated summaries. 25 respondents indicated that the summaries were good, accurate, or complete, feeling they effectively captured the essence of the videos. However, many participants noted areas for improvement: 22 responses mentioned that the summaries were missing key points, lacked specific details, or needed more explanations to be fully comprehensive. Additionally, 6 respondents suggested that including code examples, especially in programming-related videos, would enhance clarity and provide practical value.

Other feedback focused on formatting and presentation improvements, with suggestions for better text formatting, grammar, and organization to enhance readability. A few participants provided negative feedback, stating that the summaries were not helpful, missed key ideas, or focused on unimportant information. Additional comments included removing unnecessary or promotional content and adjusting the summary length to match the video's depth, ensuring it is neither too brief nor overly detailed.

\section{Threats To Validity and Limitations}\label{sec:Threats To Validity}

We discuss potential threats to the validity of our study and limitations affecting the generalizability and accuracy of our findings, focusing on internal, external, and construct validity in both the empirical evaluation and the user study.

Internal validity could be influenced by any inherent potential biases in the dataset curated from TechTube's replication package. These biases might stem from the types of programming tasks, query formulations, or video selections, potentially affecting our results. Excluding some unavailable videos likely does not significantly alter topic balance due to the dataset's diversity. In the user study, we replicated TechTube's code from their replication package to fit with our backend tool, implementing the search engine for our user study. This mitigated errors in the replicated approach. To mitigate participant bias when evaluating search results or summaries, participants were unaware of their sources, and we randomized their order to minimize placement bias. Dividing tasks into phases helped reduce participant fatigue.

External validity is limited by the dataset's scope—86 videos and 98 queries focused on programming tasks—which may not generalize to other technical domains or more complex tasks. In the user study, focusing on Python programming tasks and involving primarily computer science students may limit the applicability of our findings to other languages, domains, or user groups.

Construct validity threats in the empirical evaluation arise from using specific metrics like Hit@k, MRR, and MAP, which assess performance based on retrieving a single relevant video per query. This may not fully capture the system's ability to handle complex queries requiring multiple relevant videos. Metrics relying on relevance judgments could also introduce variability. In the user study, the subjective nature of participant evaluations might influence results. We minimized bias by randomizing the placement of search results and summaries and using a diverse set of Python tasks to reduce topic-specific familiarity.

A notable limitation of our approach is handling complex queries that require retrieving multiple non-continuous relevant segments from a video. Currently, the system identifies and highlights one continuous segment based on cosine similarity and sequential proximity, which may miss important non-adjacent segments. Addressing this limitation would enhance the system's ability to provide information spanning multiple distinct sections of a video.

\section{Acknowledgments}
This work was supported in part by the US National Science Foundation grant 1846142. Ahmad Tayeb is sponsored in part by the Saudi Arabian Cultural Mission (SACM) and King Abdulaziz University (KAU).
\section{Conclusions}
\label{sec:Conclusions and Future Work}

In this paper, we introduced \approachSIM, a method for retrieving video tutorials and identifying their most relevant fragments using Transformer-based models. Our approach generates semantically rich embeddings of video transcripts to accurately match user queries with video content, re-ranking segmented transcript chunks to pinpoint the most pertinent video segments. Additionally, we incorporate GPT-4-generated summaries to help users quickly assess video content. Through our empirical evaluation and user study, we demonstrated that \approachSIM outperforms the baseline method, TechTube, in retrieving both relevant videos and their most relevant fragments. Our user study revealed a strong developer preference for AI summaries over original video descriptions.

\bibliographystyle{plain}

\bibliography{references}

\end{document}